# Sustaining high-solar-activity research

Heliophysics Decadal Survey 2024–2033 white paper

Lead Author:
Albert Y. Shih, NASA Goddard Space Flight Center
albert.y.shih@nasa.gov; 0000-0001-6874-2594

Co-Authors:
Amir Caspi, Southwest Research Institute; amir@boulder.swri.edu; 0000-0001-8702-8273
Jessie Duncan, NASA Postdoctoral Program / NASA GSFC; dunca369@umn.edu; 0000-0002-6872-4406
Lindsay Glesener, University of Minnesota; glesener@umn.edu; 0000-0001-7092-2703
Silvina E. Guidoni, American University / NASA GSFC; guidoni@american.edu; 0000-0003-1439-4218
Katharine K. Reeves, Center for Astrophysics | Harvard & Smithsonian; kreeves@cfa.harvard.edu; 0000-0002-6903-6832

Synopsis:
Research efforts that require observations of high solar activity, such as multiwavelength studies of large solar flares and CMEs, must contend with the 11-year solar cycle to a degree unparalleled by other segments of heliophysics.  While the "fallow" years around each solar minimum can be a great time frame to build the next major solar observatory, the corresponding funding opportunity and any preceding technology developments would need to be strategically timed.  Even then, it can be challenging for scientists on soft money to continue ongoing research efforts instead of switching to other, more consistent topics.  The maximum of solar cycle 25 is particularly concerning due to the lack of a US-led major mission targeting high solar activity, which could result in significant attrition of expertise in the field.  We recommend the development of a strategic program of missions and analysis that ensures optimal science return for each solar maximum while sustaining the research community between maxima.

# Background

The 11-year solar cycle is one of the most fundamental aspects and drivers of heliophysics, and observations during the peaks of solar activity provide the most detail into, and hence the fullest understanding of, the processes of impulsive energy release and transport at the Sun. However, research efforts that target high solar activity, such as multiwavelength studies of large solar flares and CMEs, must contend with the solar cycle to a degree unparalleled by other segments of heliophysics. In contrast to the wealth of observations that ought to be available around each solar maximum, high-solar-activity researchers on soft money can find it challenging to continue ongoing research efforts during the "fallow" years around each solar minimum, rather than switching to other, more consistent topics.

The *RHESSI* SMEX mission, which operated from 2002 to 2018 (a 14-year extended mission beyond its two-year prime mission), provided an unusual level of stability in the high-solar-activity research community despite the variation of the solar cycle. *RHESSI* made transformational observations of particle acceleration and plasma heating in solar flares [1]. With its longevity, which spanned the solar minimum between solar cycles 23 and 24, *RHESSI* helped the research community thrive. While some scientists were directly paid by *RHESSI*, most scientists obtained their own funding for research efforts by leveraging the continuous generation of *RHESSI* data.

Unfortunately, we are entering a period of uncertain health and potential peril for this community. Not only has the *RHESSI* mission ended, there is no US-led successor to *RHESSI* planned for solar cycle 25. Neither of the two major solar-flare observatories that had been proposed for the upcoming maximum in ~2025 – the *FOXSI* SMEX mission concept [2] and the *FIERCE* MIDEX mission concept [3] – were selected. This observational gap in the Heliophysics System Observatory (HSO) is partially mitigated by missions expected to be in operation from space agencies in Europe and China, but those missions do not obviate the need for advanced observatories such as *FOXSI* or *FIERCE*. In addition, the Solar Flare Energy Release (SolFER) DRIVE Science Center [4] was not selected for Phase II, which would have supported continuing, substantial flare research separate from funding through a major mission.

# Recommendation

We recommend that NASA develop a strategic program of missions and analysis that ensures optimal science return for each solar maximum while sustaining the research community between maxima. This program would be founded on an openly developed roadmap for NASA Heliophysics. Failing to take full advantage of a solar maximum should be considered a significant missed opportunity, analogous to failing to take full advantage of a total solar eclipse, except solar maxima are even rarer. This strategic program has three primary facets:

- **Timing:** There must be at least one mission announcement of opportunity (AO) that is strategically timed to be well aligned with an upcoming solar maximum, and mission concepts should be evaluated and prioritized based on how well they take advantage of the solar maximum. The timing of the AO and subsequent process should be driven by the strategic goal – i.e., making scientific breakthroughs during the next maximum – rather than letting annual budgets or other government bureaucracy cause delays in implementing a mission and consequently negatively impact the achievable science. The AO timeline should include margin for common delay amounts as determined by analyzing historical performance data, as well as for extreme events such as government shutdowns.

- In the current paradigm, even if the AO is originally intended to be aligned with the next solar maximum, the accumulation of delays at each stage of the proposal process can significantly affect the alignment (Figure 1). Proposal teams often feel compelled to propose to <u>every</u> opportunity in the hope that the timing will happen to work out for one of the proposals. Due to the competitive nature of AOs, proposal teams will feel compelled to put in as much effort as possible rather than being satisfied with a "selectable" proposal. This frequent proposing, even to opportunities that overlap in time, comes at the expense of traditional research, and thus can cause irreparable damage to the careers of early-career scientists. On the other hand, proposers are reluctant to propose missions truly optimized for high solar activity, for fear of incurring a major weakness in the proposal evaluation if NASA timelines slip.

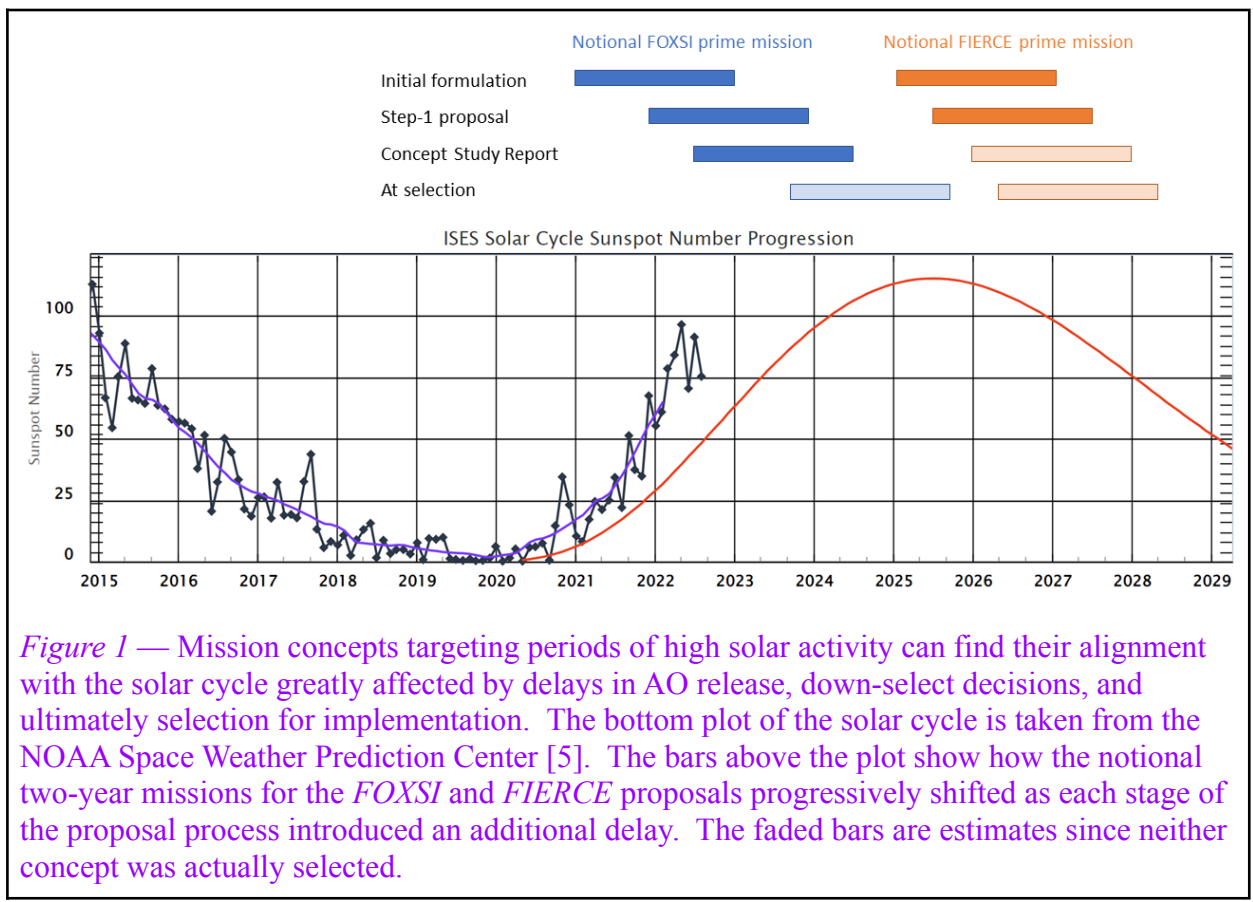

*Figure 1* — Mission concepts targeting periods of high solar activity can find their alignment with the solar cycle greatly affected by delays in AO release, down-select decisions, and ultimately selection for implementation. The bottom plot of the solar cycle is taken from the NOAA Space Weather Prediction Center [5]. The bars above the plot show how the notional two-year missions for the *FOXSI* and *FIERCE* proposals progressively shifted as each stage of the proposal process introduced an additional delay. The faded bars are estimates since neither concept was actually selected.

- **Coordination:** When possible, mission AOs should be coordinated with other mission opportunities – either from NASA or from other agencies – to facilitate the achievement of system-level science by prescribing the science topics for some of these opportunities. In the absence of a flagship mission (e.g., [6]), some science investigations require several next-generation instruments that must instead be divided across multiple smaller missions that have to be separately funded. An overarching plan for system-level science, including ensuring overlap (e.g., at least one year) between complementary major observatories, would enable the co-optimization of instruments across missions to achieve shared science objectives not possible with any individual mission.

    ○ In the current paradigm, each mission concept cannot rely on the existence of other not-yet-funded missions when proposing science objectives, which leads to each mission not being able to make the strongest case for its total science return or potentially flying duplicative instrumentation. The achievement of multiple-mission science objectives in the HSO has been largely through serendipity, although some such objectives can be proposed through the Senior Review process (but only for missions that are already funded).

- **Sustainment:** There must be an express objective to maintain the health of the US community that performs research associated with high solar activity. Sufficient funding should be allocated through multiple avenues to ensure that soft-money scientists can reasonably be retained between solar maxima. Technology development – both instrumentation and modeling – should be strategically funded with timing and cadence to support the timed mission AOs.

    ○ In the current paradigm, the sustainment of such research implicitly depends on the existence of a major mission and the knock-on effects of the continuous generation of data by that mission. The community suffers attrition when there is no such major mission. The restriction of the Heliophysics Guest Investigators solicitation to currently operating Heliophysics missions is particularly problematic. This deters the use of legacy mission data and also deters the use of data from non-Heliophysics missions (e.g., astrophysics missions that are able to observe the Sun, such as *NuSTAR*).

While some of the above facets are already supported through programmatic decisions made internally within NASA HQ, the lack of transparently communicated priorities means that there is significant "wasted" effort by the community as a result of excessive proposing, especially when a proposal concept may turn out to be a "non-starter" from the outset due to programmatic factors. The recommended strategic program would operate in a transparent manner with accountability. The communication out to the community would allow for scientists and institutions to more efficiently plan their resources and avoid burn-out. Consequently, there would be more efficient use of NASA resources, as missions targeting high solar activity would be proposed at the most beneficial time and achieve the greatest science return.